\documentclass[12pt]{article}

\usepackage{graphicx}
\begin{document}

\begin{center}
{\bf AdS black holes in the framework of nonlinear electrodynamics, thermodynamics and Joule–-Thomson expansion } \\
\vspace{5mm} S. I. Kruglov
\footnote{E-mail: serguei.krouglov@utoronto.ca}
\underline{}
\vspace{3mm}

\textit{Department of Physics, University of Toronto, \\60 St. Georges St.,
Toronto, ON M5S 1A7, Canada\\
Department of Chemical and Physical Sciences, University of Toronto,\\
3359 Mississauga Road North, Mississauga, ON L5L 1C6, Canada \\
Canadian Quantum Research Center, 204-3002 32 Ave Vernon, BC V1T 2L7, Canada}\\

\vspace{5mm}
\end{center}
\begin{abstract}
Thermodynamics and phase transitions of magnetic Anti-de Sitter black holes are studied. We consider an extended phase space thermodynamics with the cosmological constant being a thermodynamic pressure and the black hole mass is treated as a chemical enthalpy. The extended phase space thermodynamics of black holes mimics the behaviour of the Van der Walls liquid. Quantities conjugated to the coupling of nonlinear electrodynamics (NED) and a magnetic charge are obtained. Thermodynamic critical points of phase transitions are investigated. It is demonstrated that the first law of black hole thermodynamics and the generalized Smarr relation take place. The Joule--Thomson adiabatic expansion of NED-AdS black holes is studied. The dependence of inversion temperature on pressure and the minimum of inversion temperature are found.
\end{abstract}

\section{Introduction}

Black hole thermodynamics is important for understanding and developing quantum gravity. The matter is that the microscopic structure
of the black hole is unknown. By now, there are evidences that black holes behave as a thermodynamics system \cite{Bardeen,Jacobson, Padmanabhan}.  The area of the black hole is considered as the entropy and the surface gravity is connected with the temperature \cite{Bekenstein,Hawking}. The interest in AdS space-time is increased because of a holographic picture where black holes are systems which are dual to conformal field theories \cite{Maldacena,Witten,Witten1}. In addition, space-time with a negative cosmological constant allows phase transitions in black holes \cite{Page}. The holography helps to solve some problems in quantum chromodynamics \cite{Kovtun} and condensed matter physics  \cite{Kovtun1,Hartnoll}. The cosmological constant in gravity theory is understood as a vacuum
expectation value of fields, and therefore, it may vary. Hence, the cosmological constant should be included in black hole thermodynamics.
In extended phase space thermodynamics of black holes the cosmological constant is treated as a thermodynamics pressure which is a conjugate to a volume and phase transitions mimic Van der Waals liquid-gas behaviour \cite{Dolan}, \cite{Teo}.

In this paper, we study nonlinear electrodynamics (NED), proposed in \cite{Kr0,Kr1,Kr2}, coupled to AdS-gravity in extended phase space thermodynamics. In some NED \cite{Born,Kr3,Kr4,Kr5} (and others) singularities in centers of point-like electric charges are absent. In addition, quantum corrections are taken into account in Euler--Heisenberg NED \cite{Heisenberg}. Born--Infeld-AdS gravity was studied in \cite{Mann1,Hendi1,Zou, Fernando, Dey,Cai,Fernando1,Myung,Banerjee,Miskovic} where  an analogy with Van der Waals fluids was shown. $P-V$ criticality and extended phase thermodynamics were studied in \cite{Teo,Mann1,Majhi,Jafarzade,Pradhan,Pradhan1,Bhattacharya,Cong,Mann3,Mann2}.

Here, the Joule--Thomson adiabatic thermal expansion of NED-AdS black holes with heating and cooling regimes is studied.  The black hole mass is considered as the enthalpy, that is constant during the Joule--Thomson expansion.

The structure of the paper is as follows. We obtain the NED-AdS metric function of magnetic black holes and corrections to the Reissner--Nordstr\"{o}m solution in section 2. In section 3, the first law of black hole thermodynamics in the extended phase space with a negative cosmological constant, being a pressure, is studied. We define the thermodynamic magnetic potential and the thermodynamic conjugate to the NED coupling. It is shown that the generalized Smarr relation holds. The critical specific volume, critical temperature and critical pressure are found in section 4. In subsection 4.1 the Gibbs free energy is analysed showing the critical behaviour. The Joule--Thomson adiabatic NED-AdS black hole expansion is studied in section 5. Inversion temperature and pressure are obtained and their dependence on model parameters is investigated. Section 6 is a summary.

We use units with $c=1$, $\hbar=1$, $k_B=1$.

\section{NED-AdS black hole solution}

The action of NED coupled to Einstein-AdS gravity is given by
\begin{equation}
I=\int d^{4}x\sqrt{-g}\left(\frac{R-2\Lambda}{16\pi G_N}+\mathcal{L}(\mathcal{F}) \right),
\label{1}
\end{equation}
where $G_N$ is Newton's constant, $\Lambda=-3/l^2$ is the negative cosmological constant and $l$ is the AdS radius. Here, the NED Lagrangian, proposed in \cite{Kr0} (see also \cite{Kr1}, \cite{Kr2}), is
\begin{equation}
{\cal L}(\mathcal{F}) =-\frac{{\cal F}}{1+\sqrt[4]{2|\beta{\cal F}|}},
\label{2}
\end{equation}
where ${\cal F}=F^{\mu\nu}F_{\mu\nu}/4=(B^2-E^2)/2$ is the field strength invariant, $B$ and $E$ are the magnetic and electric fields, respectively, $F_{\mu\nu}=\partial_\mu A_\nu-\partial_\nu A_\mu$. At $\beta=0$ Eq. (2) is converted into Maxwell's electrodynamics Lagrangian.
The reason for choosing the theory with the NED Lagrangian (2) is in its simplicity. The metric and other functions are simple elementary functions. The gravitation and electromagnetic field equations are obtained by variation of action (1) with respect to metric $g_{\mu\nu}$ and potential $A_\mu$
\begin{equation}
R_{\mu\nu}-\frac{1}{2}g_{\mu \nu}R+\Lambda g_{\mu \nu} =8\pi G_N T_{\mu \nu},
\label{3}
 \end{equation}
\begin{equation}
\partial _{\mu }\left( \sqrt{-g}\mathcal{L}_{\mathcal{F}}F^{\mu \nu}\right)=0,
\label{4}
\end{equation}
with $\mathcal{L}_{\mathcal{F}}=\partial \mathcal{L}( \mathcal{F})/\partial \mathcal{F}$.
Making use of Eq, (2) we obtain the energy-momentum tensor of NED
\begin{equation}
 T_{\mu\nu }=F_{\mu\rho }F_{\nu }^{~\rho }\mathcal{L}_{\mathcal{F}}+g_{\mu \nu }\mathcal{L}\left( \mathcal{F}\right).
\label{5}
\end{equation}
Equations (3)-(5) take place for a general function ${\cal L}(\mathcal{F})$.
Here, we consider space-time with the spherical symmetry,
\begin{equation}
ds^{2}=-f(r)dt^{2}+\frac{1}{f(r)}dr^{2}+r^{2}\left( d\theta
^{2}+\sin ^{2}\theta d\phi ^{2}\right).
\label{6}
\end{equation}
The metric function $f(r)$ is given by \cite{Bronnikov}
\begin{equation}
f(r)=1-\frac{2m(r)G_N}{r},
\label{7}
\end{equation}
with the mass function
\begin{equation}
m(r)=M+\int_{\infty}^{r}\rho (r)r^{2}dr.
\label{8}
\end{equation}
The $M$ is the ADM black hole mass and $\rho$ is the energy density. We study magnetic black holes because models with the electrically charged black holes, with the Maxwell weak-field limit, have singularities \cite{Bronnikov}. Then, $B=q/r^2$ and the Lorentz invariant is $\mathcal{F}=q^2/(2r^4)$, where $q$ is the magnetic charge. Therefore, we consider black holes as magnetic monopoles.
By virtue of Eq. (5) the energy density with the cosmological constant term is given by
\begin{equation}
\rho=\frac{q^2}{2r^3(r+\sqrt{q}\beta^{1/4})}-\frac{3}{2G_Nl^2}.
\label{9}
\end{equation}
This and following equations are obtained for NED Lagrangian (2). The mass function $m(r)$ follows from Eqs. (8) and (9)
\begin{equation}
m(r)=M+\frac{q^{3/2}}{2\beta^{1/4}}\ln\left(\frac{r}{r+\sqrt{q}\beta^{1/4}}\right)-\frac{r^3}{2G_Nl^2}.
\label{10}
\end{equation}
It's worth mentioning that in our NED model (2) the total magnetic mass $m_M=\int_0^\infty \rho r^2 dr$ is infinite. From Eqs. (7) and (10) one finds the metric function
\begin{equation}
f(r)=1-\frac{2MG_N}{r}-\frac{q^{3/2}G_N}{\beta^{1/4}r}\ln\left(\frac{r}{r+\sqrt{q}\beta^{1/4}}\right)+\frac{r^2}{l^2}.
\label{11}
\end{equation}
With the help of Eq. (11), ignoring the cosmological constant ($l\rightarrow \infty$), we obtain the metric function when $r$ approaches to infinity
\begin{equation}
f(r)=1-\frac{2MG_N}{r}+\frac{q^2G_N}{r^2}-\frac{q^{5/2}\beta^{1/4}G_N}{2r^3}+\frac{q^3\sqrt{\beta}G_N}{3r^4}+\mathcal{O}(r^{-5}).
\label{12}
\end{equation}
Making use of Eq. (12), at $\beta=0$, one finds the metric function of the Reissner--Nordstr\"{o}m space-time.
From Eq. (12) one can obtain corrections to the Reissner--Nordstr\"{o}m black holes in the order of $\mathcal{O}(r^{-3})$.
In Fig. 1, the plots of metric function (11) are depicted at $G_N=1$, $M=1$ for different parameters.
\begin{figure}[h]
\includegraphics  {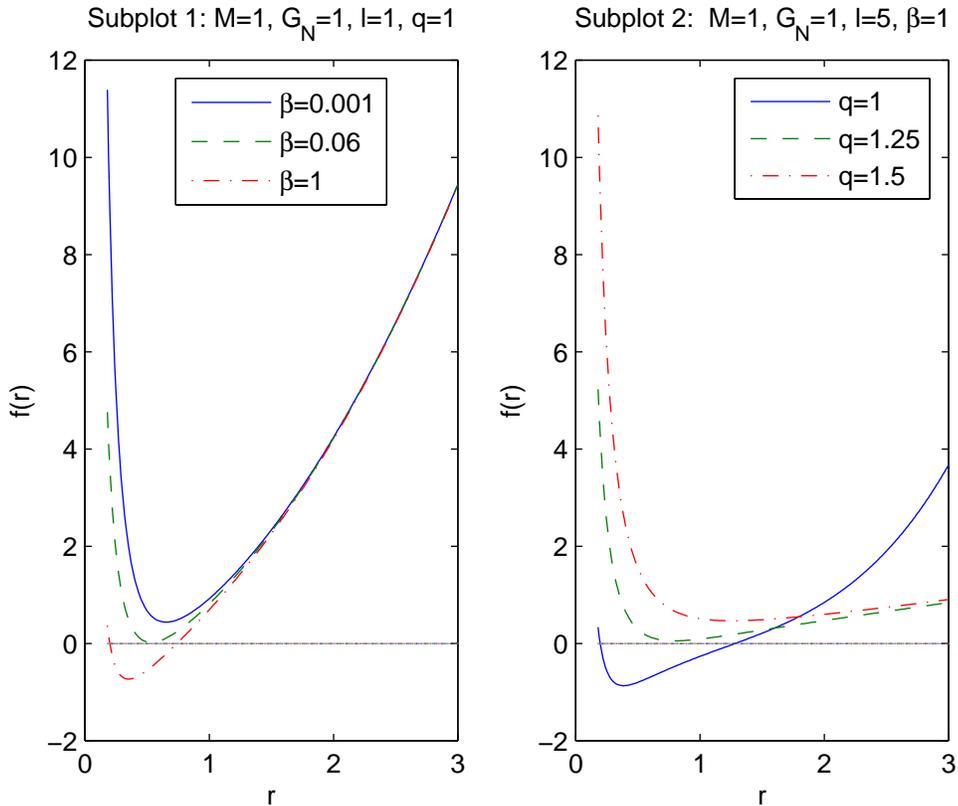}
\caption{\label{fig.1} The plots of the function $f(r)$ at $G_N=1$, $M=1$.}
\end{figure}
According to Fig. 1 black holes could have two event horizon radii or one extreme event horizon radius for various parameters. It should be noted that when there is no horizon, it is not a black hole, and there is a naked singularity. Figure 1, subplot 1, shows that when the coupling $\beta$ is increasing, with fixed magnetic charge $q$ and the AdS radius $l$, the event horizon radius $r_+$ (the largest root of equation $f(r_+)=0$) increases. But, if magnetic charge increases, at fixed NED parameter $\beta$ and the cosmological constant, the event horizon radius decreases.

\section{The first law of black hole thermodynamics and Smarr relation}

Introducing the pressure which is connected with a negative cosmological constant $\Lambda$ \cite{Kastor,Dolan1,Cvetic}, we can formulate the generalized first law of black hole thermodynamics: $dM=TdS+ VdP+\Omega dJ+\Phi dq$. Here, $M$ is treated as a chemical enthalpy \cite{Kastor} so that $M=U+PV$, where $U$ is the internal energy.
Making use of the Euler scaling argument \cite{Smarr,Kastor} we can find the generalized Smarr relation from the first law of BH thermodynamics. The dimensional analysis, with $G_N=1$, gives $[M]=L$, $[S]=L^2$, $[P]=L^{-2}$, $[J]=L^2$, $[q]=L$, $[\beta]=L^2$. Here, $\beta$ is considered as a thermodynamic variable. By using the Euler theorem, one obtains the mass
\begin{equation}
M=2S\frac{\partial M}{\partial S}-2P\frac{\partial M}{\partial P}+2J\frac{\partial M}{\partial J}+q\frac{\partial M}{\partial q}+2\beta\frac{\partial M}{\partial \beta},
\label{13}
\end{equation}
where the $\partial M/\partial \beta\equiv {\cal B}$ is the thermodynamic conjugate to NED coupling $\beta$.  The black hole entropy $S$, volume $V$ and pressure $P$ are given by \cite{Myers,Myers1}
\begin{equation}
S=\pi r_+^2,~~~V=\frac{4}{3}\pi r_+^3,~~~P=-\frac{\Lambda}{8\pi}=\frac{3}{8\pi l^2}.
\label{14}
\end{equation}
With the aid of Eq. (11) and the equation for the event horizon radius $f(r_+)=0$ we obtain the black hole mass
\begin{equation}
M=\frac{r_+}{2G_N}+\frac{r_+^3}{2G_Nl^2}-\frac{q^{3/2}}{2\beta^{1/4}}\ln\left(\frac{r_+}{r_++\sqrt{q}\beta^{1/4}}\right).
\label{15}
\end{equation}
Making the limit $\beta\rightarrow 0$ in Eq. (15), one finds the mass function of Maxwell-AdS black holes
\begin{equation}
M_l=\frac{r_+}{2G_N}+\frac{r_+^3}{2G_Nl^2}+\frac{ q^2}{2r_+}.
\label{16}
\end{equation}
By virtue of Eq. (15), for non-rotating black hole, $J=0$, with $G_N=1$, we obtain
\[
dM=\left(\frac{1}{2}+\frac{3r_+^2}{2l^2}-\frac{q^2}{2r_+(r_++\beta^{1/4}\sqrt{q})}\right)dr_+
-\frac{r_+^3}{l^3}dl
\]
\[
+\left(-\frac{3\sqrt{q}}{4\beta^{1/4}}\ln\left(\frac{r_+}{r_++\sqrt{q}\beta^{1/4}}\right)+
\frac{q}{4(r_++\beta^{1/4}\sqrt{q})}\right)dq
\]
\begin{equation}
+\left(\frac{q^{3/2}}{8\beta^{5/4}}\ln\left(\frac{r_+}{r_++\sqrt{q}\beta^{1/4}}\right)+
\frac{q^2}{8\beta(r_++\beta^{1/4}\sqrt{q})}\right)d\beta.
\label{17}
\end{equation}
The Hawking temperature is given by
\begin{equation}
T=\frac{f'(r)|_{r=r_+}}{4\pi},
\label{18}
\end{equation}
with $f'(r)=\partial f(r)/\partial r$.
From Eqs. (11), (18) and $f(r_+)=0$ at $G_N=1$, one finds
\begin{equation}
T=\frac{1}{4\pi}\biggl(\frac{1}{r_+}+\frac{3r_+}{l^2}-\frac{q^2}{r_+^2(r_++ \beta^{1/4}\sqrt{q})}\biggr).
\label{19}
\end{equation}
Making use of Eqs. (14), (17) and (19) we obtain the first law of black hole thermodynamics
\begin{equation}
dM = TdS + VdP + \Phi dq + {\cal B}d\beta.
\label{20}
\end{equation}
From Eqs. (17) and (20) one finds the magnetic potential $\Phi$ and vacuum polarization ${\cal B}$
\[
\Phi =-\frac{3\sqrt{q}}{4\beta^{1/4}}\ln\left(\frac{r_+}{r_++\sqrt{q}\beta^{1/4}}\right)+
\frac{q}{4(r_++\beta^{1/4}\sqrt{q})},
\]
\begin{equation}
{\cal B}=\frac{q^{3/2}}{8\beta^{5/4}}\ln\left(\frac{r_+}{r_++\sqrt{q}\beta^{1/4}}\right)+
\frac{q^2}{8\beta(r_++\beta^{1/4}\sqrt{q})}.
\label{21}
\end{equation}
It should be noted that $\lim_{\beta\rightarrow 0}\Phi=q/r_+$, and potential $\Phi$ at $\beta=0$ corresponds to the  potential of point-like magnetic monopole $q/r_+$. Thus, the coupling $\beta$ smoothes the singularity at $r_+=0$. The plots of potential $\Phi$ and ${\cal B}$ versus $r_+$ are depicted in Fig. 2.
\begin{figure}[h]
\includegraphics {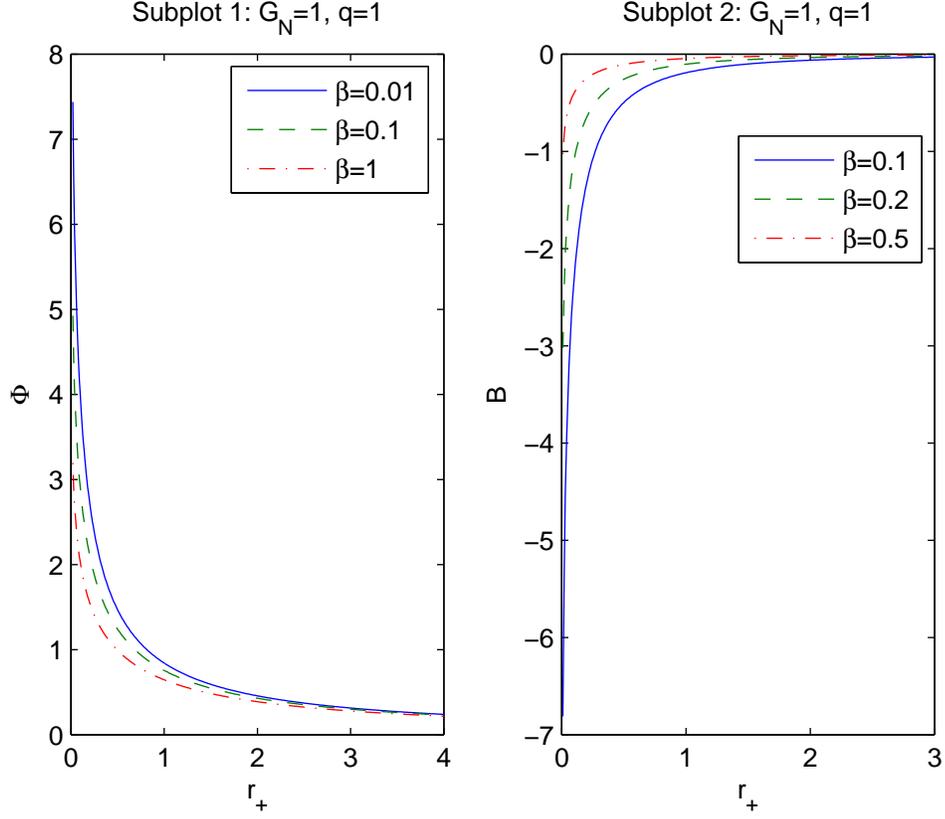}
\caption{\label{fig.2} The plots of the functions $\Phi$ and ${\cal B}$ vs. $r_+$ at $q=1$.}
\end{figure}
Figure 2, subplot 1, shows that when coupling $\beta$ increases the magnetic potential $\Phi$  is decreasing and at $r_+\rightarrow \infty$ it vanishes ($\Phi(\infty)=0$).
In accordance with Fig. 2, subplot 2, when parameter $\beta$ is increasing vacuum polarization $|{\cal B}|$ decreases and as $r_+\rightarrow \infty$ we have ${\cal B}(\infty)=0$.
Making use of Eqs. (14), (15), (19) and (21), one can verify that the generalized Smarr relation holds
\begin{equation}
M=2ST-2PV+q\Phi+2\beta{\cal B}.
\label{22}
\end{equation}

\section{Black hole thermodynamics}

Making use of Eq. (19) we obtain the black hole equation of state (EoS)
\begin{equation}
P=\frac{T}{2r_+}-\frac{1}{8\pi r_+^2}+\frac{q^2}{8\pi r_+^3(r_++\beta^{1/4}\sqrt{q})}.
\label{23}
\end{equation}
At $\beta=0$, Eq. (23) becomes EoS for a charged Maxwell-AdS black hole \cite{Mann2}. EoS of charged AdS black hole is similar to EoS of the Van der Waals fluid, when specific volume $v=2l_Pr_+$ \cite{Mann2} with $l_P=\sqrt{G_N}=1$. Then Eq. (23) is rewritten as
\begin{equation}
P=\frac{T}{v}-\frac{1}{2\pi v^2}+\frac{2q^2}{\pi v^3(v+2\beta^{1/4}\sqrt{q})}.
\label{24}
\end{equation}
Critical points, which correspond to inflection points in the $P-v$ diagram, can be found by solving equations
\[
\frac{\partial P}{\partial v}=-\frac{T}{v^2}+\frac{1}{\pi v^3}-\frac{4q^2(2v+3\beta^{1/4}\sqrt{q})}{\pi v^4(v+2\beta^{1/4}\sqrt{q})^2}=0,
\]
\begin{equation}
\frac{\partial^2 P}{\partial v^2}=\frac{2T}{v^3}-\frac{3}{\pi v^4}+\frac{8q^2(12q\sqrt{\beta}+15v\beta^{1/4}\sqrt{q}+5v^2)}{\pi v^5(v+2 \beta^{1/4}\sqrt{q})^3}=0.
\label{25}
\end{equation}
Making use of Eq. (25) one finds the equation for the critical points
\begin{equation}
8q^2\left(3v_c^2+6q\sqrt{\beta}+8\beta^{1/4}\sqrt{q}v_c\right)-v_c\left(v_c+2\beta^{1/4}\sqrt{q}\right)^3=0.
\label{26}
\end{equation}
By virtue of Eq. (25) we obtain the critical temperature and pressure
\begin{equation}
T_c=\frac{1}{\pi v_c}-\frac{4q^2\left(2v_c+3\beta^{1/4}\sqrt{q}\right)}{\pi v_c^2\left(v_c+2\beta^{1/4}\sqrt{q}\right)^2},
\label{27}
\end{equation}
\begin{equation}
P_c=\frac{1}{2\pi v_c^2}-\frac{2q^2\left(3v_c+4\beta^{1/4}\sqrt{q}\right)}{\pi v_c^3\left(v_c+2\beta^{1/4}\sqrt{q}\right)^2}.
\label{28}
\end{equation}
Solutions (approximate) to Eq. (26), critical temperatures and pressures are presented in Table 1.
\begin{table}[ht]
\caption{Critical values of the specific volume and temperature at $q_m=1$}
\centering
\begin{tabular}{c c c c c c c c c c}\\[1ex]
\hline
$\beta$ & 0.1 & 0.2 & 0.3 & 0.4 & 0.5 & 0.6 & 0.7 & 0.8 & 0.9  \\[0.5ex]
\hline
$v_{c}$ &4.021 & 3.870& 3.772 & 3.697 & 3.637 & 3.585 & 3.541& 3.501& 3.466 \\[0.5ex]
\hline
$T_{c}$ &0.0502 & 0.0517 & 0.0527 & 0.0535 & 0.0542 & 0.0548 & 0.0553 & 0.0558 & 0.0562 \\[0.5ex]
\hline
$P_{c}$ &0.0046 & 0.0048 & 0.0050 & 0.0052 & 0.0054 & 0.0055 & 0.0056 & 0.0057 & 0.0058 \\[0.5ex]
\hline
\end{tabular}
\end{table}
The plots of $P-v$ diagrams for various parameters $\beta$ are depicted in Fig. 3.
\begin{figure}[h]
\includegraphics {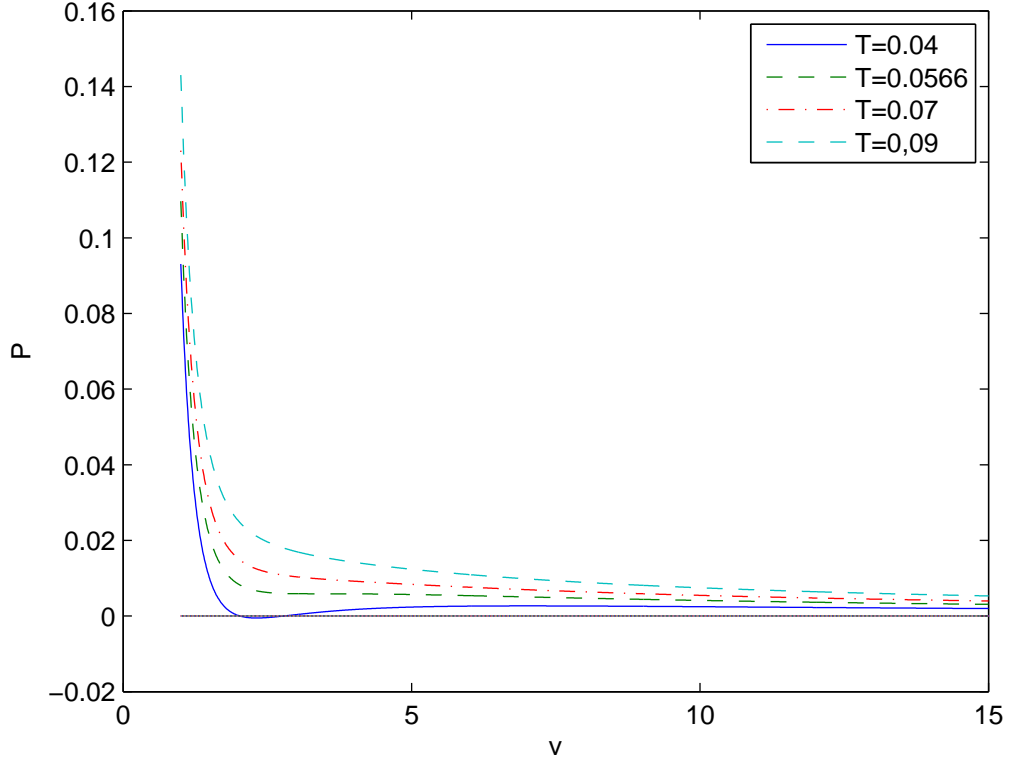}
\caption{\label{fig.3} The plots of the function $P$ vs. $v$ at $q=1$, $\beta=1$. The critical isotherm corresponds to $T_{c}=0.0566$.}
\end{figure}
Approximate solution to Eq. (26) at $q=1$, $\beta=1$ is $v_{c}\approx 3.433$ and the critical temperature and pressure are $T_{c}=0.0566$, and $P_c=0.0059$, respectively. The critical isotherm corresponds to $T_{c}=0.0566$ but non-critical behaviour of $P-v$ diagrams are for $T=0.4$, $0.7$ and $0.9$. Two upper curves in Fig. 3 show one-phase state for $T > T_c$ and correspond to the ``ideal gas". The lower solid line shows a two-phase behavior for $T < T_c$. From Eqs. (27) and (28) we obtain the critical ratio
\begin{equation}
\rho_c=\frac{P_cv_c}{T_c}=\frac{v_c(v_c+2\beta^{1/4}\sqrt{q})^2-4q^2(3v_c+4\beta^{1/4}\sqrt{q})}{2[v_c(v_c+2\beta^{1/4}\sqrt{q})^2
-4q^2(2v_c+3\beta^{1/4}\sqrt{q})]}.
\label{29}
\end{equation}
By virtue of Eq. (29), at $\beta=0$ ($v_c^2=24q^2$), one finds the ratio $\rho_c=3/8$ that corresponds to a Van der Waals fluid.

\subsection{ The Gibbs free energy}

The Gibbs free energy is given by
\begin{equation}
G=M-TS,
\label{30}
\end{equation}
where the black hole mass $M$ is treated as a chemical enthalpy.  From Eqs. (20) and (30) one finds $dG=VdP + \Phi dq + {\cal B}d\beta-SdT$. Therefore, $G$ is stationary at fixed ($P,q,\beta,T$).
With the help of Eqs. (15), (19) and (30), ($G_N=1$) we obtain
\begin{equation}
G=\frac{r_+}{4}-\frac{2\pi r_+^3P}{3}-\frac{q^{3/2}}{2\beta^{1/4}}\ln\left(\frac{r_+}{r_++\sqrt{q}\beta^{1/4}}\right)+\frac{q^2}{4(r_++\beta^{1/4}\sqrt{q})}.
\label{31}
\end{equation}
We will study the dependence of the Gibbs free energy $G$ on the temperature $T$ for different fixed ($P,q,\beta$) values.
The plots of $G$ versus $T$, where we took onto account that $r_+$ is a function of $P$ and $T$ (see Eq. (23)), are depicted in Fig. 4.
\begin{figure}[h]
\includegraphics {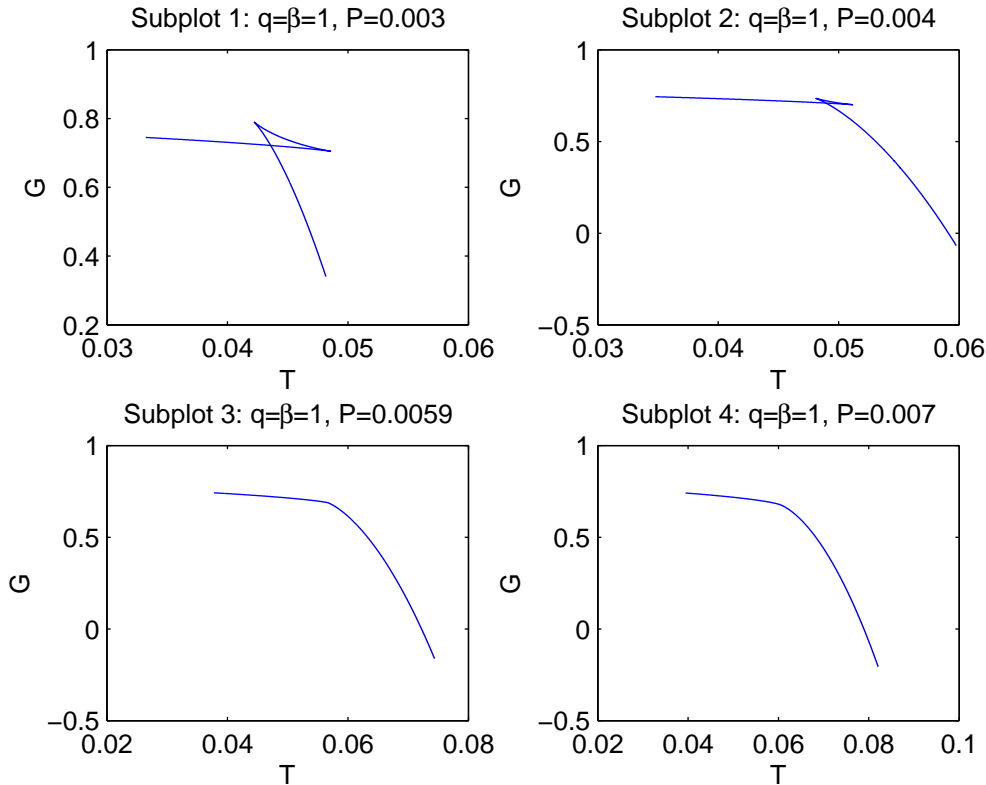}
\caption{\label{fig.4} The plots of the Gibbs free energy $G$ vs. $T$ with $q=1$, $\beta=1$.}
\end{figure}
Subplots 1 and 2 in Fig. 4 show first-order phase transitions between small and large black holes. The Gibbs free energy at $P<P_c$ has 'swallowtail' behaviour. Points on the curves correspond to different event horizon radius of black holes.
At the critical point, $v_c\approx 3.433$ ($r_c\approx 1.7165$), $T_c\approx 0.0566$, and $P_c\approx 0.0059$ at $q=\beta=1$ in subplot 3, second-order phase transition takes place. Subplot 3 shows that there is a cusp at the critical point. Hance, the second derivative of Gibbs free energy $G$ with respect to temperature $T$ is discontinuous at that point. We check this analytically in Appendix. The behaviour is similar to that of the Reissner–-Nordstr\"{o}m-AdS black hole. In subplot 4 for $P>P_c$ only a single phase exists, and we have the smooth single-valued curve.
The plots of entropy $S$ versus temperature $T$ at $q=\beta=1$ were depicted in Fig. 5. According to Fig. 5, subplots 1 and 2, entropy is ambiguous function of the temperature for some region. This confirms that for this region first-order phase transitions occur. Figure 5, subplot 3 shows the second-order phase transition, where the tangent of the slope in the critical point $r_c\approx 1.72$, $P_c\approx0.0059$, $T_c\approx 0.0566$  is infinite. The critical point separates a low-entropy state and a high-entropy state. In accordance with Fig. 5, subplot 4, there is not a critical behaviour of a black hole.
\begin{figure}[h]
\includegraphics {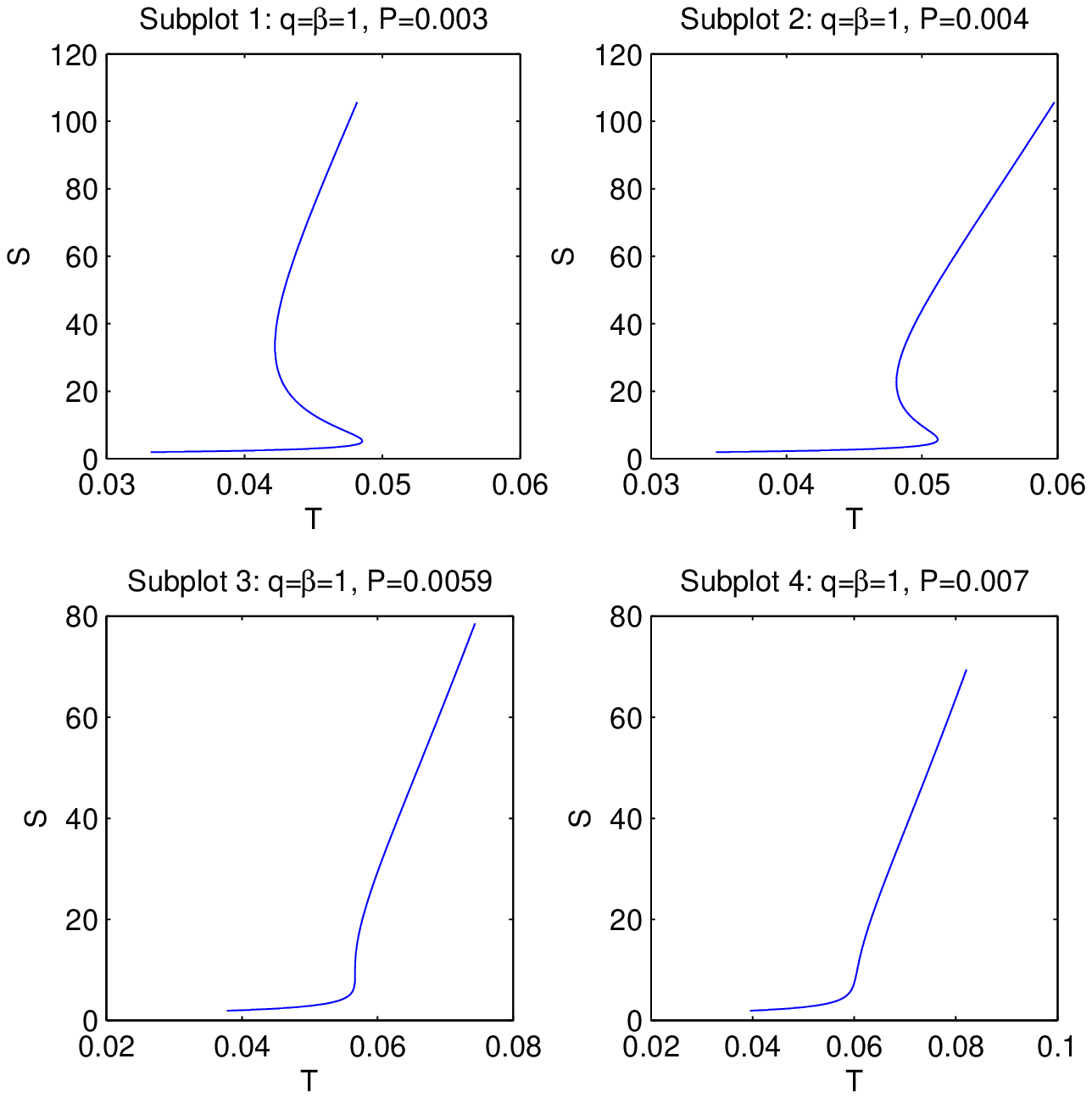}
\caption{\label{fig.5} The plots of entropy $S$ vs. temperature $T$ at $q=1$, $\beta=1$.}
\end{figure}

\section{Joule--Thomson expansion of black holes}

During the Joule--Thomson adiabatic expansion (isenthalpic) the black hole mass $M$, that is the enthalpy, is constant. The Joule--Thomson thermodynamic coefficient, which characterises cooling-heating phases, is given by
\begin{equation}
\mu_J=\left(\frac{\partial T}{\partial P}\right)_M=\frac{1}{C_P}\left[ T\left(\frac{\partial V}{\partial T}\right)_P-V\right]=\frac{(\partial T/\partial r_+)_M}{(\partial P/\partial r_+)_M}.
\label{32}
\end{equation}
The Joule--Thomson coefficient $\mu_J$, according to Eq. (32), is the slope of the tangent in $P-T$ diagrams.
At the inversion temperature $T_i$, when $\mu_J(T_i)=0$, the sign of $\mu_J$ is changed. When the initial temperature during the isenthalpic expansion is higher than inversion temperature $T_i$, the temperature decreases that is the cooling phase ($\mu_J>0$), however if the initial temperature is lower than $T_i$, the final temperature increases for this  heating phase. Making use of Eq. (32) and equation $\mu_J(T_i)=0$ one obtains the inversion temperature
\begin{equation}
T_i=V\left(\frac{\partial T}{\partial V}\right)_P=\frac{r_+}{3}\left(\frac{\partial T}{\partial r_+}\right)_P.
\label{33}
\end{equation}
In fact, the inversion temperature is a borderline between cooling and heating processes. The inversion temperature
goes through the maxima of $P-T$ diagrams and the slope of $P-T$ diagrams is changed and separating cooling and heating phases
of black holes \cite{Yaraie,Rizwan}. Black hole EoS (23) can be represented as
\begin{equation}
T=\frac{1}{4\pi r_+}+2P r_+-\frac{q^2}{4\pi r_+^2(r_++\beta^{1/4}\sqrt{q})}.
\label{34}
\end{equation}
At $\beta=0$ Eq. (34) is converted into the Maxwell-AdS black hole EoS.
From  Eq. (15) and equation $P=3/(8\pi l^2)$ we obtain
\begin{equation}
P=\frac{3}{4\pi r_+^3}\left[M-\frac{r_+}{2}+\frac{q^{3/2}}{2\beta^{1/4}}\ln\left(\frac{r_+}{r_++\sqrt{q}\beta^{1/4}}\right)\right].
\label{35}
\end{equation}
Equations (34) and (35) represent the $P-T$ diagrams in the parametric form which are depicted in Fig. 6.
\begin{figure}[h]
\includegraphics {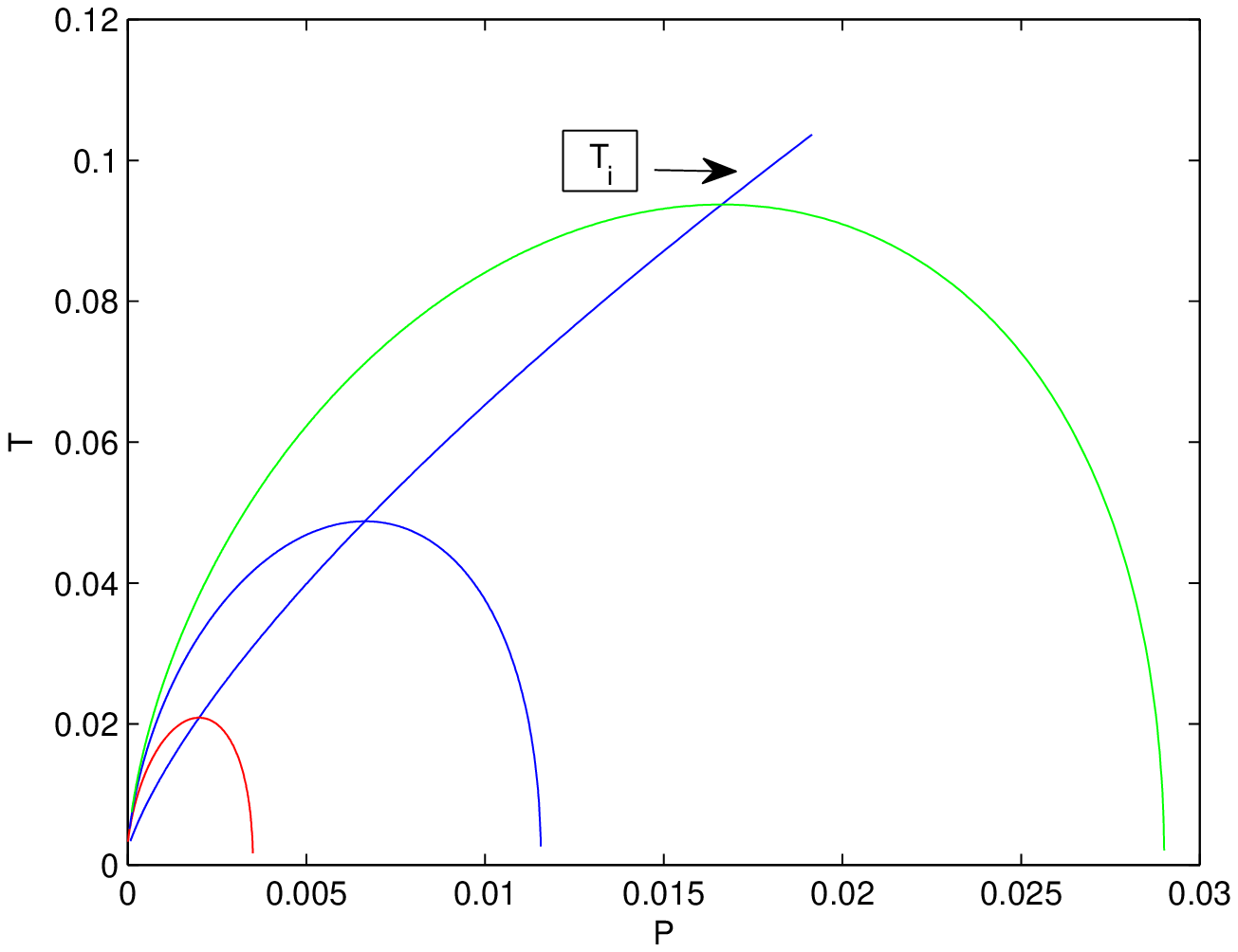}
\caption{\label{fig.6} The plots of the temperature $T$ vs. pressure $P$  and the inversion temperature $T_i$ with $q=10$, $\beta=1$.
The $P-T$ diagrams correspond to masses $M=12$, $14$, $16$ from bottom to top.}
\end{figure}
In accordance with Fig. 6 when the black hole mass is increasing the inversion point increases.
Making use of Eqs. (33) and (34) one finds the equation for inversion pressure $P_i$
\begin{equation}
P_i=\frac{q^2\left(6r_++5\beta^{1/4}\sqrt{q}\right)}{16\pi r_+^3\left(r_++\beta^{1/4}\sqrt{q}\right)^2}-\frac{1}{4\pi r_+^2}.
\label{36}
\end{equation}
By using Eqs. (34) and (36) we obtain the inversion temperature
\begin{equation}
T_i=\frac{q^2\left(4r_++3\beta^{1/4}\sqrt{q}\right)}{8\pi r_+^2\left(r_++\beta^{1/4}\sqrt{q}\right)^2}-\frac{1}{4\pi r_+}.
\label{37}
\end{equation}
Equations (36) and (37) are equations for inversion temperature $T_i$ versus $P_i$ in the parametric form. The plots of $T_i$ versus $P_i$ are depicted in Fig. 7. According to Fig. 7, subplot 1, the inversion temperature increases with $q$.  Fig. 7, subplot 2, shows that the inversion temperature decreases when coupling $\beta$ is increasing.
\begin{figure}[h]
\includegraphics {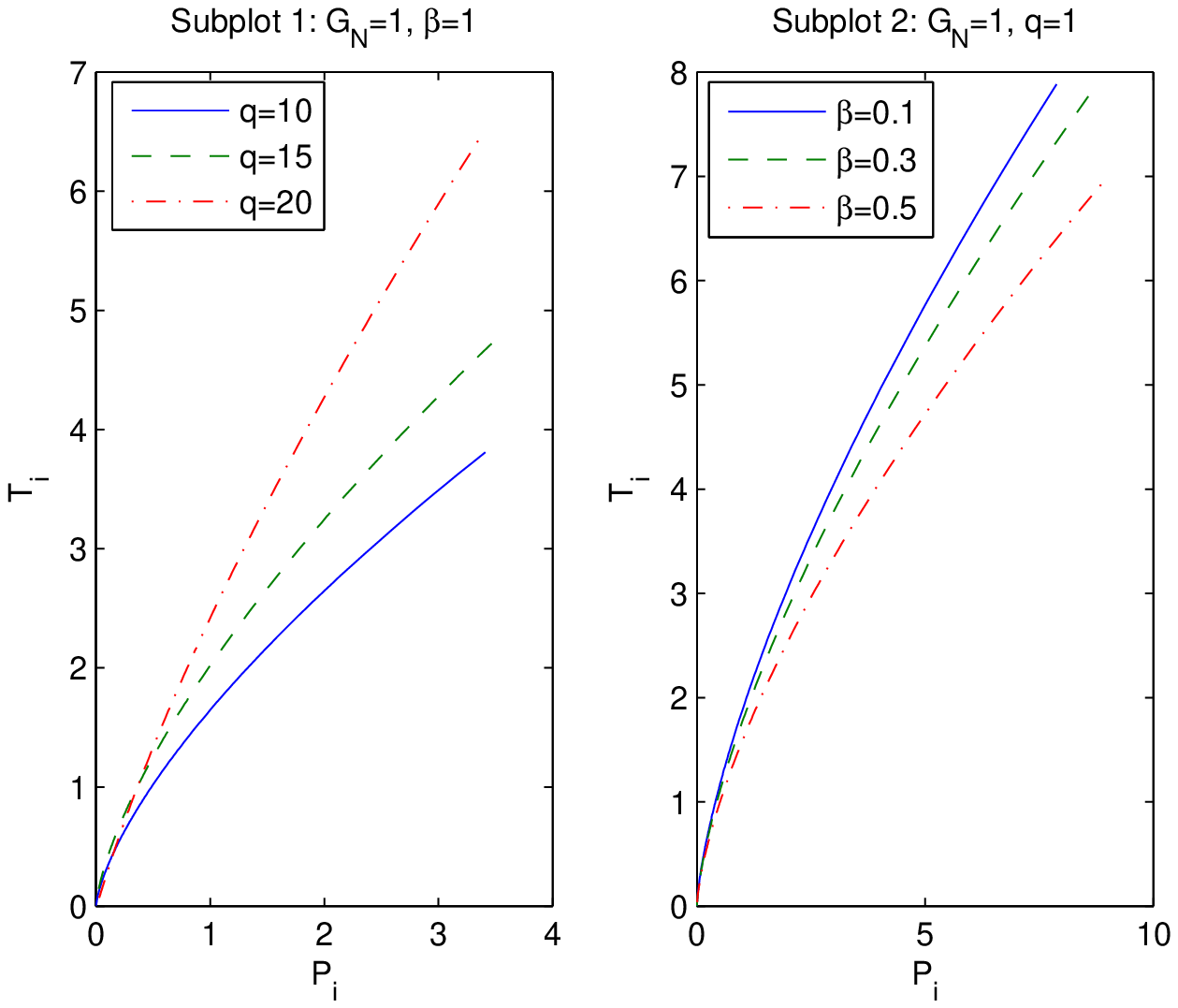}
\caption{\label{fig.7} The plots of the inversion temperature $T_i$ vs. pressure $P_i$.}
\end{figure}
If $\mu_J>0$ we have a  cooling process, while when $\mu_J<0$, a heating process occurs. The area with $\mu_J>0$, in Fig. 6, belongs to the left side of inversion temperature borderline however $\mu_J<0$ corresponds to the right site of borderline $T_i$.
Putting $P_i=0$ in Eq. (36), we obtain the equation for the minimum of event horizon radius $r_{min}$
\begin{equation}
4r_{min}^3+8br_{min}^2+2(2b^2-3q^2)r_{min}-5q^2b=0,
\label{38}
\end{equation}
where $b=\sqrt{q}\beta^{1/4}$. The discriminant of cubic Eq. (38) is negative, and therefore, Eq. (38) possesses three real solutions, one is positive physical solution $r_{min}\geq 0$ and two are negative non-physical solutions, $r_{min}<0$. The positive physical solution to Eq. (38) is given by
\[
r_{min}=2\sqrt{p}\cos\left(\frac{1}{3}\arccos\left(\frac{g}{p^{3/2}}\right)\right)-\frac{2b}{3},
\]
\begin{equation}
p=\frac{b^2}{9}+\frac{q^2}{2},~~~~g=\frac{b^3}{27}+\frac{q^2b}{8}.
\label{39}
\end{equation}
Making use of Eqs. (37) and (39), the plot of the minimum inversion temperature $T_{min}$ versus coupling $\beta$ is depicted in Fig. 8.
\begin{figure}[h]
\includegraphics {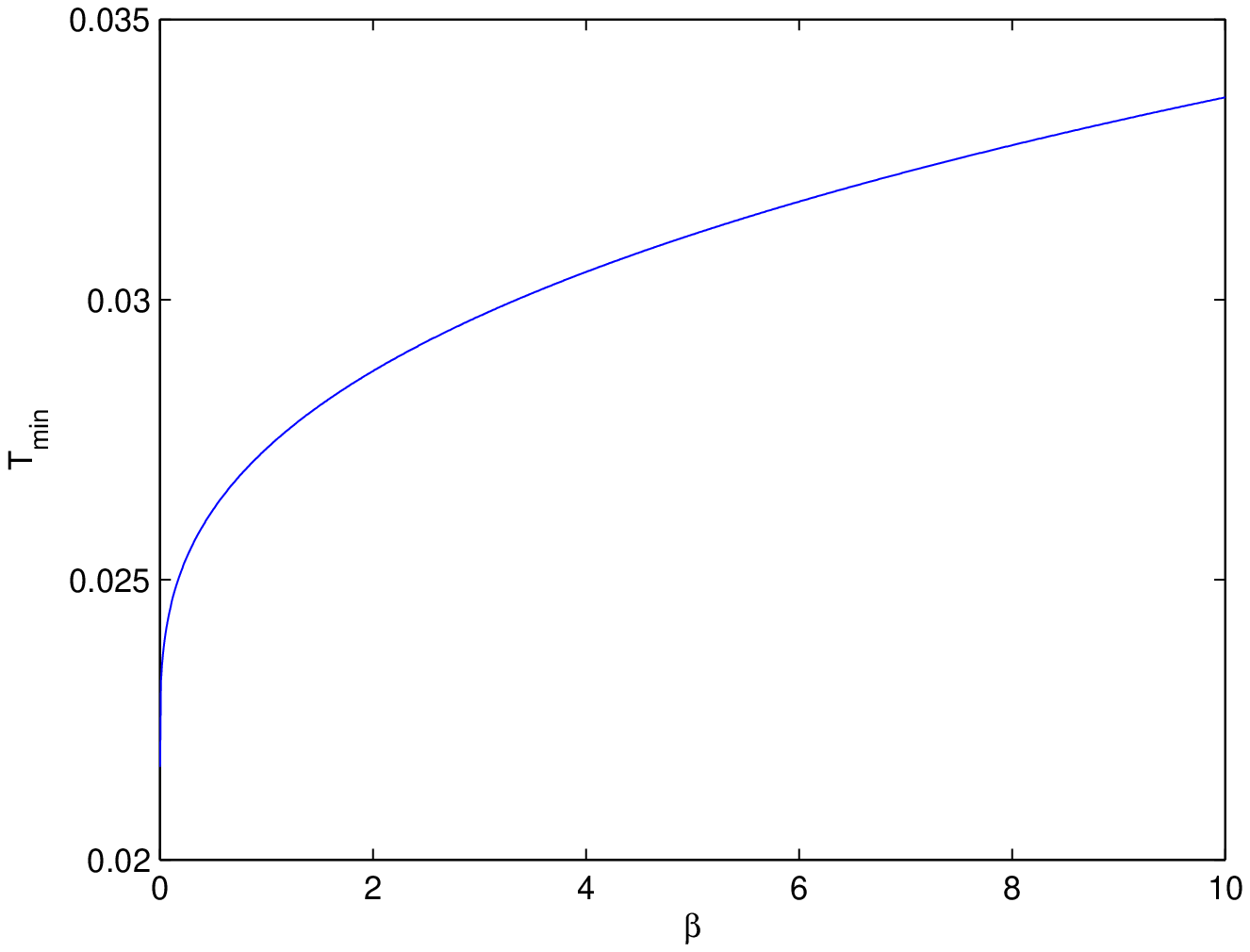}
\caption{\label{fig.8} The plots of the minimum inversion temperature $T_{min}$ vs. $\beta$ for $q=1$.}
\end{figure}
By virtue of Eqs. (38) and (39) at $\beta=0$ we obtain the minimum of inversion temperature for Maxwell-AdS magnetic black holes
\begin{equation}
T_{min}=\frac{1}{6\sqrt{6}\pi q}, ~~~~r_{min}=\frac{\sqrt{6}q}{2}.
\label{40}
\end{equation}
From Eqs. (26) and (27) at $\beta=0$ ($v_c=2\sqrt{6}q$) and Eq. (40), one finds the relation $T_{min}=T_c/2$ which is valid for electrically charged AdS black holes \cite{Aydiner}.

\section{Summary}

We have studied NED coupled to Einstein-AdS  black holes and obtained metric and mass functions. Black holes can have one or two horizons depending on magnetic charge $q$, coupling $\beta$ and the AdS radius $l$. At $l=\infty$ and $r\rightarrow\infty$ corrections to the Reissner--Nordstr\"{o}m solution are in the order of ${\cal O}(r^{-3})$. We formulated NED-AdS black hole thermodynamics in an extended thermodynamic phase space where the cosmological constant is considered as a thermodynamic pressure and the mass of the black hole is treated as the chemical enthalpy. The vacuum polarization ${\cal B}$ conjugated to NED parameter $\beta$ and thermodynamic potential $\Phi$ conjugated to magnetic charge $q$ have been obtained. It was demonstrated that the first law of black hole thermodynamics and the generalized Smarr relation take place. NED-AdS black hole thermodynamics mimics the Van der Walls liquid–gas behaviour. The critical temperature, pressure  and Gibbs free energy have been calculated and we showed that first and second-order phase transitions occur at some parameters. We have been obtained the critical ratio $\rho_c=3/8+{\cal O}(\beta)$ showing corrections to the Van der Waals critical ratio $3/8$. Similar behaviour takes place in another model of NED-AdS black hole \cite{Kr6}.

The the Joule--Thomson adiabatic isenthalpic expansion of NED-AdS black holes has been studied. The cooling and heating phase transitions occur depending on model parameters. We have obtained the inversion temperature and pressure, as functions of magnetic charge and NED coupling of black holes, which define the borderline between cooling and heating phases. The radii $r_{min}$ corresponding to minimum of inversion temperature and pressure were found. As a particular case we obtained the relation $T_{min}=T_c/2$ at $\beta=0$ connecting the critical temperature $T_c$ with the minimum inversion temperature. Previously this formula was obtained for Einstein-AdS black holes \cite{Aydiner}.

\vspace{3mm}
\textbf{Appendix}
\vspace{3mm}

One can calculate the second derivative of Gibbs free energy $G$ with respect to temperature $T$ by the relation
\begin{equation}
\frac{\partial^2 G}{\partial T^2}=\frac{\partial^2 G/\partial r_+^2}{\partial^2 T/\partial r_+^2}.
\label{41}
\end{equation}
It follows from Eq. (41) that the second derivative of $G$ with respect to $T$ is discontinuous when $\partial^2 T/\partial r_+^2=0$. Making use of Eq. (34) we obtain the second derivative of temperature at the constant pressure
\begin{equation}
\frac{\partial^2 T}{\partial r_+^2}=\frac{r_+\left(r_++\beta^{1/4}\sqrt{q}\right)^3-q^2\left(6r_+^2+8\beta^{1/4}\sqrt{q}r_++3q\sqrt{\beta}\right)}{2\pi r_+^4\left(r_++\beta^{1/4}\sqrt{q}\right)^3}.
\label{42}
\end{equation}
From Eq. (42) we obtain the equation corresponding to $\partial^2 T/\partial r_+^2=0$ giving the second derivative of $G$ with respect to $T$ being infinite
\begin{equation}
r_+\left(r_++\beta^{1/4}\sqrt{q}\right)^3-q^2\left(6r_+^2+8\beta^{1/4}\sqrt{q}r_++3q\sqrt{\beta}\right)=0.
\label{43}
\end{equation}
The approximate real and positive solution to Eq. (43) at $q=\beta=1$ is the critical event horizon radius $r_+=r_c\approx 1.7165$ corresponding to the second-order phase transition represented at Fig. 4, subplot 3. One can verify that approximate real and positive solutions to Eq. (43) for various $\beta$ are in accordance with Table 1.  This confirms that the second-order phase transition takes place when the second derivative of $G$ with respect to $T$ is discontinuous.

\end{document}